# Hyperbolic meteors: is CNEOS 2014-01-08 interstellar?

*J. Vaubaillon* [1]

In 2019 a claim was made that the CNEOS 2014-01-08 meteor is interstellar. However, apparent interstellar meteors have been detected for decades. Moreover, they are expected from any meteor observation survey, as a natural consequence of measurement error propagation. Here we examine if enough scientific data were published to identify the orbital and physical nature of CNEOS 2014-01-08. Given the lack of proof regarding the accuracy of the observation, the derivation of the trajectory, velocity and tensile strength, and given the current state of meteor observations and reduction tools, we find no scientific ground to conclude about the interstellar orbit nor the physical properties of CNEOS 2014-01-08. Moreover, given the current data release of this object, to find any piece at the bottom of the ocean seems extremely unlikely.

## 1 Introduction

At the time this article is written (late 2022), several announcements claiming that the meteor CNEOS 2014-01-08 observed by "US sensors" (likely those described in Brownet al., 1996) is of interstellar origin (Siraj & Loeb, 2019; Siraj & Loeb, 2022). However, the scientific community has not validated this claim, mainly for lack of scientific evidence. Only an analyst at the US Department of Defense declared that the accuracy of the data is good enough to confirm the interstellar origin. However, the data acquisition and reduction process are not described, preventing anyone from building on such a claim.

The goal of this article is first to provide the reader with the basics of meteor science, and to examine if enough scientific proof are available to derive a hyperbolic nature of the CNEOS 2014-01-08 object. This work is based on a web page published in late Aug. 2022[a], and is intended to be accessible to non-meteor specialists as well as the general public. We do not exclude that future scientific proof regarding CNEOS 2014-01-08 may change the conclusions presented here.

## 2 Introduction to meteor science and optical observations

### 2.1 Basics of meteor phenomena

When a meteoroid (small rock in the interplanetary space) enters the Earth atmosphere, it hits the air molecules (Ceplecha et al., 1998). The energy involved in this process is so high that atoms are excited. The disexcitement process involves the emission of light, (often) detectable by the naked eye as a meteor (a.k.a. a "shooting star"). Usual meteoroid top of the atmosphere velocity is comprised between 11 km/s (caused by Earth gravity) and $\sim 72$ km/s (the heliocentric hyperbolic limit). Meteors usually start at an altitude of $\sim 100(\pm 20)$ km. The end point depends on the size and velocity of the initial meteoroid. Meteorite falls are associated with bright meteors (fireballs). As a rule of thumb, if the initial meteoroid is large and slow enough, and is still visible at "low" altitude (typically $< 30$ km), the possibility of a meteorite fall is real. However, a fireball does not necessarily lead to meteorite fall. Any "fast" meteoroid (typically $> 40$ km/s) has nearly zero chance to survive the atmosphere entry, simply because the shock with the atmosphere is too violent, making the ablation and fragmentation process extremely efficient. In short: the fastest the meteoroid is, the less likely it is to survive.

### 2.2 Optical meteor Observations

Current and widely spread method to observe the meteors is to set up at least 2 cameras, pointing towards a similar portion of the atmosphere, able to detect the same meteor from two different sites which are at least 30 km away from each other (Koten et al., 2019). This allows, after thorough and careful calculations, the 3D-trajectory to be computed. From this trajectory, provided the data are time-resolved, the velocity can be derived.

### 2.3 Velocity

Deriving a meteor velocity and quantifying the velocity accuracy is difficult. The reasons are numerous. First, any position measurement has errors. If one computes the velocity from 2 consecutive positions, the spread in the velocity will necessarily be extremely high. The reason is simply because $v = d/t$, with $d$ being the distance between 2 positions: any position measurement error will induce a high velocity error. Another reason is that the geometry is different from one camera to the other. Suppose the meteor appear above camera 1: this camera will detect the meteor before camera 2 because camera 2 is farther away from the meteor. Suppose now that the meteor is travelling towards camera 2: instead of detecting a moving object, the camera will see a single point which increases and decreases in brightness, making deriving a 3D trajectory impossible, as well as the velocity. Regarding software data analysis, recent studies (see Section 2.4) have shown that the velocity accuracy directly depends on the method used to compute it (Egal et al., 2017; Vida et al., 2018). Finally, remember that the meteoroid velocity decreases along its path, because of the friction with the atmosphere molecules.

---

[1]IMCCE, Observatoire de Paris, univ. PSL, CNRS, Sorbonne Université, Univ. Lille, France.
Email: `jeremie.vaubaillon@obspm.fr`

[a]https://www.imcce.fr/recherche/equipes/pegase/hypermeteor

Long story short: the derivation of a meteor velocity is an extremely complicated process that required 2 PhDs in the recent years to tackle (see Egal et al. 2017; Vida et al. 2018). Most accurate instruments, dedicated to meteor velocity measurements provide at best, depending on the meteor and camera geometry (see previous paragraph), m.s$^{-1}$ accuracy velocity (Borovička et al., 2007; Egal et al., 2017; Vida et al., 2021). Several tens of m.s$^{-1}$ accuracy is expected from lower resolution instruments (time and space) (SonotaCo, 2009; Kornoš et al., 2014; Colas et al., 2020). Remember also that the meteoroid is constantly decelerating during its flight in the atmosphere (see Section 2.1).

### 2.4 From Velocity to Orbit

The derivation of a meteoroid orbit from its derived velocity is another complicated problem. This can be done analytically or numerically, and a comparison by (Clark & Wiegert, 2011) between the 2 methods shows that, in most cases, the analytical approach is correct.

However, when publishing an orbit (usually as Keplerian osculating elements), the epoch for which these elements are provided matters. Indeed, because of gravitational perturbations, any orbit in interplanetary space (slightly but) constantly changes. In particular, for an apparent hyperbolic meteor, its past orbit might simply have been perturbed by Jupiter (known to kick small bodies out of the Solar System): this does not mean it comes from another planetary system (see also e.g. Wiegert, 2014).

### 2.5 Hyperbolic Meteors

Hyperbolic meteors were detected long ago: a quick search shows that such meteors were detected several decades ago (Almond et al., 1951; Almond et al., 1952; Hajduková et al., 2019). In other words, CNEOS 2014-01-08 is not the first (apparent) interstellar meteoroid.

Even more, hyperbolic meteors are expected from any observation set: given the meteor position and velocity uncertainty (see above), any meteor orbit survey program (for a review, see Koten et al., 2019) will produce data containing apparent hyperbolic meteors. This is expected and perfectly normal, and is a natural consequence of measurement errors and error propagation given the method used to compute the velocity. The percentage of apparently hyperbolic meteors in a meteor orbit database is actually an indicator of the real accuracy a network (hardware, method + reduction pipeline) provides (Hajdukova, 1994; Hajduková et al., 2017).

Another reason is the extremely fine line between a high eccentricity orbit and a hyperbolic orbit: suppose a meteoroid comes from the end of the Solar System (let us say from a region close to Pluto). Its velocity as measured as a meteor in the Earth atmosphere will be "high" (typically > 30 km/s, depending on the orbital geometry): this is a simple consequence of the gravitation. Suppose now that an interstellar meteoroid coming from outside the Solar System at nearly zero velocity (at infinity): it will also show a "high" velocity. The difference between the two velocities might be less than 0.1 km/s. Again, given the uncertainties in deriving the velocity, it is expected that apparent hyperbolic meteors are measured from any meteor orbit survey. Remember also that the meteor samples the trajectory over a few dozen km, while the whole orbit might bring the meteoroid several hundred millions of kilometres from the Earth. In other words, we sample an extremely tiny portion of the meteoroid orbit, and the slightest error on the velocity makes it appear either bound or unbound (i.e. hyperbolic).

For a full review of the problem of apparently hyperbolic meteors, see e.g. Hajdukova et al. (2019) and Hajdukova et al. (2020).

## 3 Is CNEOS 2014-01-08 peculiar?

### 3.1 Orbit: is CNEOS 2014-01-08 interstellar?

In order to establish the hyperbolic (interstellar) character of CNEOS 2014-01-08, it would be necessary to provide at the very least:

- The accuracy of the measurement.

- The method used for the computation of the velocity.

The measurements were performed by the classified "U.S. Government sensors", about which little is known, except that there are visible wavelength detectors. Any such satellite likely orbit at an altitude higher than 200 km (otherwise, the orbit decay would make it unstable). Since meteors occur at 100 km of altitude, the physical distance between the sensor and the meteor is at least similar than any ground-based network. The space resolution has to be at least equivalent to that of the best meteor camera resolution (which is 1 arcsec, see Vida et al., 2021). Similarly, the frame rate must be at least comparable to ground meteor cameras, i.e. at least 25 fps (or combined with another sensor, see Brown et al., 1996). Given the performance of modern optics and sensors, to build a space satellite showing such performances is likely. In order to perform 3D measurement of the position and velocity, at least 2 satellites were used. The overall geometry matters in this case (as seen in sec. 2.3). Last but not least, the very method used to derive the velocity matters when one wants to compute an accurate meteor velocity.

Unfortunately, none of the space and time resolution, number and geometry of the detection, nor the very method used to compute the trajectory, velocity and orbit of CNEOS 2014-01-08 are provided. As a consequence, there is no proof that the data are sufficiently accurate to indicate an interstellar trajectory. There is no proof that CNEOS 2014-01-08 is interstellar: its apparently hyperbolic orbit might simply reflect the influence of measurement uncertainty, as expected from any meteor observation, and as reported for the past few decades. The real velocity error can easily be of several km.s$^{-1}$. This would challenge the results of a study based on a $\sim$ 1 km.s$^{-1}$ accuracy.

## 3.2 Nature: is CNEOS 2014-01-08 of high density?

The light curve of a meteoroid allows one to derive its mass (see Section 2.1). Spikes in the light curve allow the meteor scientists to derive its tensile strength, thanks to the $\rho V^2$ equation. However, the velocity considered here is not the initial velocity: this is the velocity at the time of the light curve spike. Usually, for "large" (dm scale) object, such spike occurs when the meteoroid is "low" in the atmosphere (typically below 50 km of altitude). By then, it has lost quite an amount of velocity, because it is now entering the densest part of the atmosphere. In order to derive a tensile strength, it is therefore necessary to consider the velocity change along the 3D-trajectory.

In order to derive the tensile strength of CNEOS 2014-01-08, it would be necessary to consider the velocity change along its trajectory, and consider its velocity at the time of the light curve spike. Unfortunately, no velocity change curve is provided by CNEOS. In all cases, the velocity of CNEOS 2014-01-08 below 30 km of altitude is necessary smaller than the initial velocity at the top of the atmosphere. As a consequence, there is no proof that CNEOS 2014-01-08 is of high density (or tensile strength).

Regarding the mass: in absence of velocity curve, the dynamic mass (or size) of CNEOS 2014-01-08 cannot be derived. The photometric mass assumes a luminous efficiency $\tau$. Given the latest estimates, a value of $\tau$ greater than 5% is unlikely (but more works are needed to constrain this parameter). Usually, comparing the dynamic and photometric mass allows scientists to put boundaries on possible mass and size of the meteoroids. An accuracy of a factor of 2 or 3 in mass estimate is perceived as a good estimate. As a consequence, given the data publicly available for CNEOS 2014-01-08, an accuracy of a factor of 10 or more in the mass estimate is plausible.

## 3.3 Is it possible to recover pieces of CNEOS 2014-01-08?

According to CNEOS, CNEOS 2014-01-08 fell near the northeast coast of Papua New Guinea, i.e. in the ocean. If the initial velocity derived by CNEOS is correct (44.8 km/s), then the chances of atmospheric survival is small, since the faster the meteoroid is, the less likely it is to survive the atmosphere entry (see Section 2.1). If the initial meteoroid experienced 3 fragmentation events above 20 km of altitude (which is likely given the measured light curve) then a significant fraction of mass was lost during each event.

Suppose a least one fragment made it to the bottom of the ocean: the probability to recover it is also very low. The search of meteorites on dry land proves to be extremely challenging: the search area is often very large, and even teams of more than 10 people may fail at finding anything, unless very accurate measurements and searches are performed or we have the best luck ever. The only case a meteorite was found at the bottom of a lake was Chelyabinsk in 2013, because the very location of the impact was perfectly known since it fell during the winter and made a hole in the ice of a frozen lake. Russian scientists did wait for the (warmer) spring to dive and recovered a 50 cm piece of meteorite.

Therefore, is seems very unlikely to recover a piece of CNEOS 2014-01-08 at the bottom of the ocean.

## 4 Conclusion

Given the lack of proof regarding the accuracy of the observation, the derivation of the trajectory, velocity and tensile strength, and given the current state of meteor observations and reduction tools, we find no scientific ground to conclude about the interstellar orbit nor the physical properties of CNEOS 2014-01-08. Moreover, given the current data release of this object, to find any piece at the bottom of the ocean seems extremely unlikely.

## Acknowledgements

The author is thankful to IMO and WGN for promoting meteor science.

## References

Almond M., Davies J. C., and Lovell A. C. B. (1952). "The velocity distribution of sporadic meteors. II". *MNRAS*, **112**, 21.

Almond M., Davies J. G., and Lovell A. C. B. (1951). "The velocity distribution of sporadic meteors. I.". *MNRAS*, **111**, 585.

Borovička J., Spurný P., and Koten P. (2007). "Atmospheric deceleration and light curves of Draconid meteors and implications for the structure of cometary dust". *A&A*, **473**, 661–672.

Brown P., Hildebrand A. R., Green D. W. E., Pagé D., Jacobs C., Revelle D., Tagliaferri E., Wacker J., and Wetmiller B. (1996). "The fall of the St-Robert meteorite". *Meteoritics & Planetary Science*, **31:4**, 502–517.

Ceplecha Z., Borovička J., Elford W. G., Revelle D. O., Hawkes R. L., Porubčan V., and Šimek M. (1998). "Meteor Phenomena and Bodies". *Space Science Reviews*, **84**, 327–471.

Clark D. L. and Wiegert P. A. (2011). "A numerical comparison with the Ceplecha analytical meteoroid orbit determination method". *Meteoritics & Planetary Science*, **46:8**, 1217–1225.

Colas F., Zanda B., Bouley S., Jeanne S., Malgoyre A., Birlan M., Blanpain C., Gattacceca J., Jorda L., Lecubin J., et al. (2020). "FRIPON: a worldwide network to track incoming meteoroids". *A&A*, **644**, A53.

Egal A., Gural P. S., Vaubaillon J., Colas F., and Thuillot W. (2017). "The challenge associated with the robust computation of meteor velocities from video and photographic records". *Icarus*, **294**, 43–57.

Hajdukova, M. J. (1994). "On the frequency of interstellar meteoroids.". *A&A*, **288**, 330–334.

Hajduková, Mária J., Koten P., Kornoš L., and Tóth J. (2017). "Meteoroid orbits from video meteors. The case of the Geminid stream". *Plan. Space Sci.*, **143**, 89–98.

Hajduková, Mária J., Sterken V., and Wiegert P. (2019). "Interstellar Meteoroids". In Ryabova G. O., Asher D. J., and Campbell-Brown M. J., editors, *Meteoroids: Sources of Meteors on Earth and Beyond*, page 235.

Hajdukova M., Sterken V., Wiegert P., and KorNoš L. (2020). "The challenge of identifying interstellar meteors". *Plan. Space Sci.*, **192**, 105060.

Korноš L., Koukal J., Piffl R., and Tóth J. (2014). "EDMOND Meteor Database". In Gyssens M., Roggemans P., and Zoladek P., editors, *Proceedings of the International Meteor Conference, Poznan, Poland, 22-25 August 2013*. pages 23–25.

Koten P., Rendtel J., Shrbený L., Gural P., Borovička J., and Kozak P. (2019). "Meteors and Meteor Showers as Observed by Optical Techniques". In Ryabova G. O., Asher D. J., and Campbell-Brown M. J., editors, *Meteoroids: Sources of Meteors on Earth and Beyond*, page 90.

Siraj A. and Loeb A. (2019). "The 2019 Discovery of a Meteor of Interstellar Origin". *arXiv e-prints*, page arXiv:1904.07224.

Siraj A. and Loeb A. (2022). "New Constraints on the Composition and Initial Speed of CNEOS 2014-01-08". *Research Notes of the American Astronomical Society*, **6:4**, 81.

SonotaCo (2009). "A meteor shower catalog based on video observations in 2007-2008". *WGN, Journal of IMO*, **37**, 55–62.

Vida D., Brown P. G., and Campbell-Brown M. (2018). "Modelling the measurement accuracy of pre-atmosphere velocities of meteoroids". *MNRAS*, **479**, 4307–4319.

Vida D., Brown P. G., Campbell-Brown M., Weryk R. J., Stober G., and McCormack J. P. (2021). "High precision meteor observations with the Canadian automated meteor observatory: Data reduction pipeline and application to meteoroid mechanical strength measurements". *Icarus*, **354**, 114097.

Wiegert P. A. (2014). "Hyperbolic meteors: Interstellar or generated locally via the gravitational slingshot effect?". *Icarus*, **242**, 112–121.